\documentstyle{article}
\title{
\bf  Percolation on a Feynman Diagram 
}
\author{ {\it D.A. Johnston}\\
         Dept. of Mathematics\\
         Heriot-Watt University\\
         Riccarton\\
         Edinburgh, EH14 4AS, Scotland\\ \\
and\\ \\
         {\it P. Plech\' a\v{c}}\\
         Mathematical Institute\\
         24-29 St Giles'\\
         Oxford\\
         OX1 3LB}
\date {9th May 1997}         
\begin{document}
  \maketitle
                      {\Large
                      \begin{abstract}
%
In a recent paper \cite{dp} we investigated Potts models
on ``thin'' random graphs -- generic Feynman diagrams, using
the idea that such models may be expressed as the $N \rightarrow 1$
limit of a matrix model. The models displayed
first order transitions
for all $q>2$, giving identical behaviour to the corresponding
Bethe lattice.

We use here one of the results of \cite{dp}
namely a general saddle point solution for a $q$ state Potts
model expressed as a function of $q$, to investigate some
peculiar features of the percolative limit $q \rightarrow 1$
and compare the results with those on the Bethe lattice.
%
                        \end{abstract} }
%
  \thispagestyle{empty}
%
%
  \newpage
%
                  \pagenumbering{arabic}

\section{Introduction and $q \ge 2$ Solutions}

In a series of papers \cite{dp,1,2,3,4}
we have exploited 
the observation made in \cite{5}
that an ensemble of random graphs
could be thought of as arising from
the perturbative Feynman diagram expansion of a scalar integral
to investigate spin models living on such graphs.
The random graphs are of unrestricted topology
as can be seen by thinking of the scalar model
as the $N \rightarrow 1$ limit of an $N \times N$
Hermitian matrix model. We consequently
denote them as ``thin'' graphs, to distinguish them from
the planar ``fat'' graphs which 
appear in the $N \rightarrow \infty$ limit familiar from two dimensional
quantum gravity.

The random graph spin models are of interest because
they display mean field behaviour due to the
tree-like local structure of the graphs \cite{6}. 
Random graphs, which are closed, have an advantage over genuine
tree-like structures such as the Bethe lattice for
both numerical and analytical work because dominant boundary effects
are absent. The numerical advantages
of random graphs over the Bethe lattice have also been noted by other authors \cite{5a}.
The equilibrium behaviour on random graphs 
of ferromagnetic Ising models \cite{2} and spin glasses
\cite{3}
was found to parallel that of the equivalent model on the appropriate
Bethe lattice with the same number of neighbours
\cite{7}.

If we take the Hamiltonian for a q-state Potts model to be
\begin{equation}
H =   \beta \sum_{<ij>} ( \delta_{\sigma_i, \sigma_j} -1),
\end{equation}
where the spins $\sigma_i$ take on $q$ values,
the partition function on
$\phi^3$ (i.e. 3-regular) random graphs with $2n$
vertices is \cite{dp}
\begin{equation}
Z_n(\beta) \times N_n = {1 \over 2 \pi i} \oint { d \lambda \over
\lambda^{2n + 1}} \int { {\prod_{i=1}^q d \phi_i  \over 2 \pi \sqrt{\det K}}
\exp (- A )},
\label{part}
\end{equation}
where the action $A$ is given by
\begin{equation}
A = { 1 \over 2 } \sum_{i=1}^{q} \phi_i^2 - c \sum_{i<j} \phi_i \phi_j -{\lambda \over 3} \sum_{i=1}^q \phi_i^3,
\label{qstate}
\end{equation}
$K$ is the inverse of the quadratic form in the action,
the coupling $c$ is related to the temperature as
\begin{equation}
c = { 1\over  ( \exp( 2 \beta)  + q-2) }
\label{coup}
\end{equation}
and $N_n$ is the number of undecorated (no spin) $\phi^3$ graphs
with $2n$ vertices
\begin{equation}
N_n = \left( {1 \over 6} \right)^{2n} { ( 6 n - 1 ) !! \over ( 2 n ) !!
}.
\end{equation}
The role of the contour integration is to pick out graphs with $2n$ vertices and the 
prefactor of $N_n$ disentangles the factorial growth in the number of graphs
themselves from any non-analyticity due to a transition
in the decorating spin model. 

The model may be solved in a saddle point approximation, valid for large $n$,
and one finds a high temperature solution of the form $\phi_i= 1 - (q-1)c, \forall i$
and low temperature broken symmetry solutions $\phi_i= \ldots \phi_{q-1} \ne \phi_q$.
There are two low temperature branches, the first being
\begin{eqnarray}
\phi_{1 \ldots q-1} &=& { 1 -  (q-3) c - \sqrt{1 - 2 (q-1) c + (q-5) (q-1) c^2} \over 2}
\nonumber \\ 
\phi_q &=& { 1 +  (q-1) c + \sqrt{ 1 - 2 (q-1) c + (q-5) (q-1) c^2} \over 2}
\label{qsols}
\end{eqnarray}
the second having the signs in front of the square roots reversed
\begin{eqnarray}
\phi_{1 \ldots q-1} &=& { 1 -  (q-3) c + \sqrt{1 - 2 (q-1) c + (q-5) (q-1) c^2} \over 2}
\nonumber \\
\phi_q &=& { 1 +  (q-1) c - \sqrt{ 1 - 2 (q-1) c + (q-5) (q-1) c^2} \over 2}.
\label{qsols2}
\end{eqnarray}

The low temperature solutions can be recovered from the saddle point equations
for an ``effective action'' $A$, namely
$\partial A / \partial \phi
= \partial A / \partial \tilde \phi = 0$
where 
\begin{eqnarray}
A = {1 \over 2} ( q - 1) \left[ 1 - c ( q - 2) \right] \phi^2  - { 1 \over 3} ( q -1) \phi^3
+ {1 \over 2} \tilde \phi^2  - {1 \over 3} \tilde \phi^3 - c ( q -1 ) \phi \tilde \phi
\label{app1}
\end{eqnarray}
with $\phi=\phi_{1 \ldots q-1}$, $\tilde \phi = \phi_q$.
Similarly, 
at high temperature
one can write another effective action
\begin{eqnarray}
A_0 = { q \over 2} (1 - c ( q - 1) ) \phi_0^2 - { q \over 3} \phi_0^3
\label{app2}
\end{eqnarray}
whose saddle point equation gives $\phi_0 = 1 - ( q - 1) c$.
Both equs.(\ref{app1},\ref{app2}) follow from imposing the expected symmetry
on the $\phi_i$ in the original action in equ.(\ref{qstate}).

The Potts magnetisation
\begin{equation}
m = { \phi_q^3 \over \left( \sum_{i=1}^{q} \phi_i^3 \right)}
\label{mag1}
\end{equation}
is related to the standard order parameter by
\begin{equation}
M = { q \max ( m ) - 1 \over q -1 }.
\end{equation}
and gives a clear picture of the topology of the phase diagram.
In Fig.1 we plot $m$ against $c$ for $q=4$, this being
representative of all the $q>2$ models. 
The upper (solid) branch is given by $m$ for the solutions 
in equ.(\ref{qsols}) where it gives a maximum and the lower
(dashed) branch is given by $m$ for the solutions 
in equ.(\ref{qsols2}).
The high temperature (dotted) line is at $m=\tilde m = 1/q$.

A first order transition occurs for $q>2$ at 
\begin{equation}
c(Q) = { 1 - ( q - 1)^{-1/3} \over q -2}.
\label{ccrit}
\end{equation}
when the low temperature 
and high temperature
saddle point effective actions become equal.
One finds a jump in the magnetisation at the first order transition point of
\begin{equation}
\Delta M = { q - 2 \over  q -1}.
\label{mj}
\end{equation}
Spinodal points are present at {\bf P}
where the high temperature solution joins the lower branch
\begin{equation}
c(P) = {1 \over 2 q - 1}
\end{equation}
and {\bf O} where the low temperature branches first become real
\begin{equation}
c(O) = { q -1 - 2 \sqrt{ q - 1} \over (q - 1 ) ( q - 5)}.
\end{equation} 
For $q=2$ {\bf O,P,Q} merge and one recovers
the continuous mean field Ising transition. All the
exponents, critical values and jumps are in agreement
with the calculations in \cite{10}
for the Bethe lattice proper.

\section{Percolation and the $q \rightarrow 1$ limit}

As we have a solution and expressions for the critical points
that are apparently valid for all $q$ we shall now attempt
to emulate the work of \cite{lmp} for the Bethe lattice
and explore the percolative limit $q \rightarrow 1$.
The solution of the Bethe lattice  percolation
problem is of course well known from elementary
considerations \cite{flory,book}. One finds that the
percolation threshold on a Bethe lattice with co-ordination
number $z$ is $p_c = 1 / (z - 1 )$ and other cluster
quantities are also explicitly calculable.
Our objective here, as in \cite{lmp} for the Bethe lattice
proper, is rather to shed light on the
properties of the Potts model correspondence with percolation.

The mapping between the Potts model and a percolation problem
was first written down by Fortuin and Kastelyn \cite{11}
and is the basis of the cluster algorithms
that have been so successful in combating critical slowing
down in simulations \cite{12}.
The q-state Potts model partition function may be written
as a bond partition function
in terms of spin clusters
\begin{equation}
Z = \sum_{configs} p^o ( 1 - p)^u q^{N_c}
\label{bond}
\end{equation}
where $o = \sharp \; occupied \; bonds$, $u = \sharp \; unoccupied \; bonds$
and $N_c = \sharp \; clusters$. The factor $p$, which converts the ``geometrical''
spin clusters into critical spin clusters is $1-\exp( - 2  \beta)$
and the sum over configurations, in distinction to standard lattices,
includes a sum over different Feynman diagrams.
From equ.(\ref{bond}) one can see that the mean number of connected clusters
per vertex $N_o$ (averaged over the ensemble of Feynman diagrams) is given by
\begin{equation}
N_o = \lim_{n \rightarrow \infty} \; \lim_{q \rightarrow 1} \;   {-\log \left( Z_n \right) \over n ( q -1 )}
\end{equation} 
where $Z_n$ is the partition function in equ.(\ref{part}).
Other thermodynamic quantities in the Potts model
are related to percolative quantities, for instance the magnetisation
$m$ to the percolation probability $P$
\begin{equation}
P = \lim_{q \rightarrow 1} \;   { q \; m - 1 \over (q-1) } ,
\label{P}
\end{equation}
and the magnetic susceptibility $\chi$ to the mean
size of finite clusters $S$
\begin{equation}
S \; ( 1 - P)  = \lim_{q \rightarrow 1} \;   { \chi \over (q-1) }.
\label{S}
\end{equation}
In principle the above relations apply per site, but 
with the uniform random graphs we consider it is possible
to consider spatially averaged values and drop any site indices. 

To get some understanding of the limit $q \rightarrow 1$ 
that is required in the above let us first
look at the region $1<q<2$. In Fig.2 we have plotted
the values of $c(O),c(P)$ and $c(Q)$ where we can see 
that they fan out again from equality at $q=2$ in the same order
as for $q>2$ (i.e. {\bf Q} is sandwiched between an upper value
at {\bf O} and a lower value at {\bf P}). We can also see
that $c(O)$ and $c(Q)$ diverge as $q \rightarrow 1$. The sign
of the jump in the magnetisation is {\it reversed}
with respect to the $q>2$ solutions as can be seen in
Fig.3, where we have taken $q=1.2$ as an illustrative
example. On reducing $c$ (i.e. $T$) a first order transition
occurs at {\bf Q} from the horizontal
high temperature solution 
to the {\it lower} dashed branch of the low temperature
curve. The portions {\bf QP} and the upper branch to the left of {\bf P}
are metastable, whereas the portion {\bf OQ} of the lower branch
and the remainder of the high temperature line to the left of {\bf P}
are unstable.

As we have noted, as $q \rightarrow 1$, $c(Q) \rightarrow \infty$
or $\beta_{crit} \rightarrow 0$. The first order transition is
therefore clearly {\it not} that associated with the percolation
problem as 
\begin{equation}
p = 1 - \exp ( -2 \beta )
\label{pprob}
\end{equation}
and the known Bethe lattice solution (for $z=3$) is $p_{crit} = 1/2$.
We must look rather at the spinodal point {\bf P}
where one has $c(P) = 1 / ( 2 q -1)$, or $\exp ( 2 \beta (P) ) = q+1$.
In the limit $q \rightarrow 1$, one thus has $\exp ( 2 \beta (P) )
\rightarrow 2$ and $p_{crit} \rightarrow 1/2$, as expected.
This result is not a surprise, given the discussion of  \cite{lmp}
for percolation on the Bethe lattice, where an exactly
analogous situation occurs. In the limit
$q \rightarrow 1$ the percolation transition 
is thus associated not with the first order transition at {\bf Q} to a stable
state, but rather to the spinodal point {\bf P} where the 
high temperature phase joins a metastable branch.

It is also possible to directly calculate various percolative quantities 
in the saddle point approximation. 
For example, to find the percolation probability $P$ we note $\phi \rightarrow c$,
$\tilde \phi \rightarrow 1$ as $q \rightarrow 1$, so we have from equs.(\ref{mag1},\ref{P})
$P = 1 - c^3$, which using the relation between $c$ and $p$ 
for $q=1$, namely $c = (1-p)/p$, gives
\begin{equation}
P = 1 - \left( { 1 - p \over p} \right)^3.
\end{equation}
This reproduces the standard result for $P$ on a Bethe lattice with 3 neighbours.
The probability of a point being connected to infinity by occupied sites
is $p P$ with our conventions.

More difficult is the calculation of the susceptibility, which requires the solution of the 
saddle point equations for the effective action
in an external field $H$
\begin{eqnarray}
A = {1 \over 2} ( q - 1) \left[ 1 - c ( q - 2) \right] \phi^2  - { 1 \over 3} ( q -1) \phi^3
+ {1 \over 2} \tilde \phi^2  - {H \over 3} \tilde \phi^3 - c ( q -1 ) \phi \tilde \phi
\label{app1h}
\end{eqnarray}
(note the $H$ in front of $\tilde \phi^3$),
as we have
\begin{equation}
\lim_{q \rightarrow 1} {\chi \over q - 1} \; =  \; \lim_{q \rightarrow 1} \lim_{H \rightarrow 1} {1 \over q - 1} {d m \over d H} \; = \; 
\lim_{q \rightarrow 1} \lim_{H \rightarrow 1} \;  \left( 3 {\phi^2 \over \tilde \phi^4 } \left[ \phi {\partial \tilde \phi \over \partial H} - \tilde \phi 
{\partial  \phi \over \partial H} \right] + {\phi^3 \over \tilde \phi^3 } \right).
\label{chiH}
\end{equation}
The first two terms in equ.(\ref{chiH}) come from the implicit dependence of the solutions on $H$
and the final term is  $\partial m / \partial H$.
In calculating this one uses the expression for the magnetisation
in the presence of a field
\begin{equation}
m = { H \tilde \phi^3 \over (q - 1 ) \phi^3 + H \tilde \phi^3}.
\end{equation}
The full solutions for $\phi, \tilde \phi$ in the presence of a field
are not particularly illuminating
so we do not reproduce them here,
but we find the simple limits
\begin{eqnarray}
\lim_{q \rightarrow 1} \lim_{H \rightarrow 1} \phi  &=&  c  \nonumber \\
\lim_{q \rightarrow 1} \lim_{H \rightarrow 1} {\partial \phi \over \partial H} &=&  {c \over c - 1} \nonumber \\
\lim_{q \rightarrow 1} \lim_{H \rightarrow 1} \tilde \phi  &=&  1  \nonumber \\
\lim_{q \rightarrow 1} \lim_{H \rightarrow 1} {\partial \tilde \phi \over \partial H} &=&  -1 
\end{eqnarray}
giving
\begin{eqnarray} 
\lim_{q \rightarrow 1} {\chi \over q - 1} \; =  \; S \; ( 1 - P ) \; = \; c^3 { 1 + 2 c \over 1 - c},
\end{eqnarray}
which again reproduces the Bethe lattice result for $S$.

\section{Discussion}

The somewhat peculiar features of the correspondence between the 
$q \rightarrow 1$ limit of the Potts model and percolation
on the Bethe lattice 
in which the spinodal point is tied
to the percolative transition are clearly preserved on Feynman diagrams.
Using the relations between Potts model observables and those associated
with the percolation problem, we obtain the standard Bethe lattice
results for quantities such as $P$ and $S$.
The inventory of results in which the 
(ferromagnetic) transitions of spin models on random graphs are identical to those
on corresponding Bethe lattices is thus extended to 
percolation and Potts models analytically continued to real values of $q$.

It is also interesting to note that the Ising
spin glass transition on random graphs,
which appears in the $k \rightarrow 0$ limit of a $k$ Ising replica model,
displays some very similar features to the percolative limit discussed here. 
In such models
the spin glass transition in the (quenched) $k \rightarrow 0$ limit is associated with  a
continuous transition which  appears $\forall k \ge 2$ between the high temperature
phase and a phase that is metastable when $k>2$ \cite{2}. The ``true'' transition
in the $k$ Ising replica model is first order for $k>2$  \cite{SP,5} and is {\it not} 
that associated with the $k \rightarrow 0$ limit. 
One might hope
to be able to analyse the multi-Ising models in a similar fashion to the Potts 
models by writing down
an effective action for various $k$ and seeing if a general formula could be extracted
to allow analytical continuation in $k$.
The superficial
similarity with the percolative limit is intriguing and certainly
merits further investigation to see if there is any deeper
connection.

\clearpage \newpage

\clearpage \newpage
\begin{figure}[htb]
\vskip 20.0truecm
\includegraphics{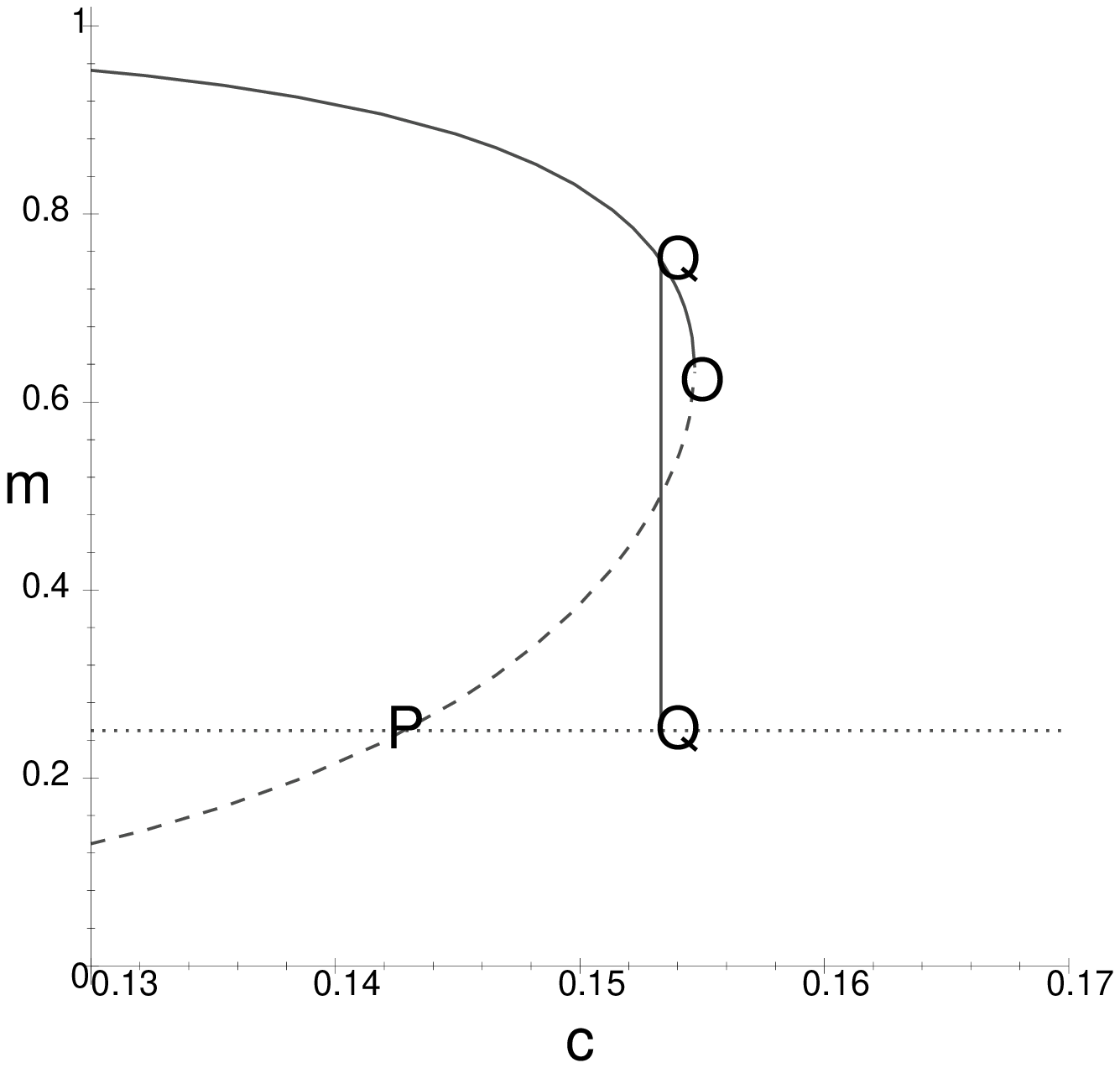}
\caption[]{\label{fig1} The magnetisation $m$
for a $4$ state Potts model as calculated from
the saddle point solutions. The high temperature branch is
shown dotted, the upper low temperature branch solid and the lower
low temperature branch dashed.}
\end{figure}
\clearpage \newpage
\begin{figure}[htb]
\vskip 20.0truecm
\includegraphics{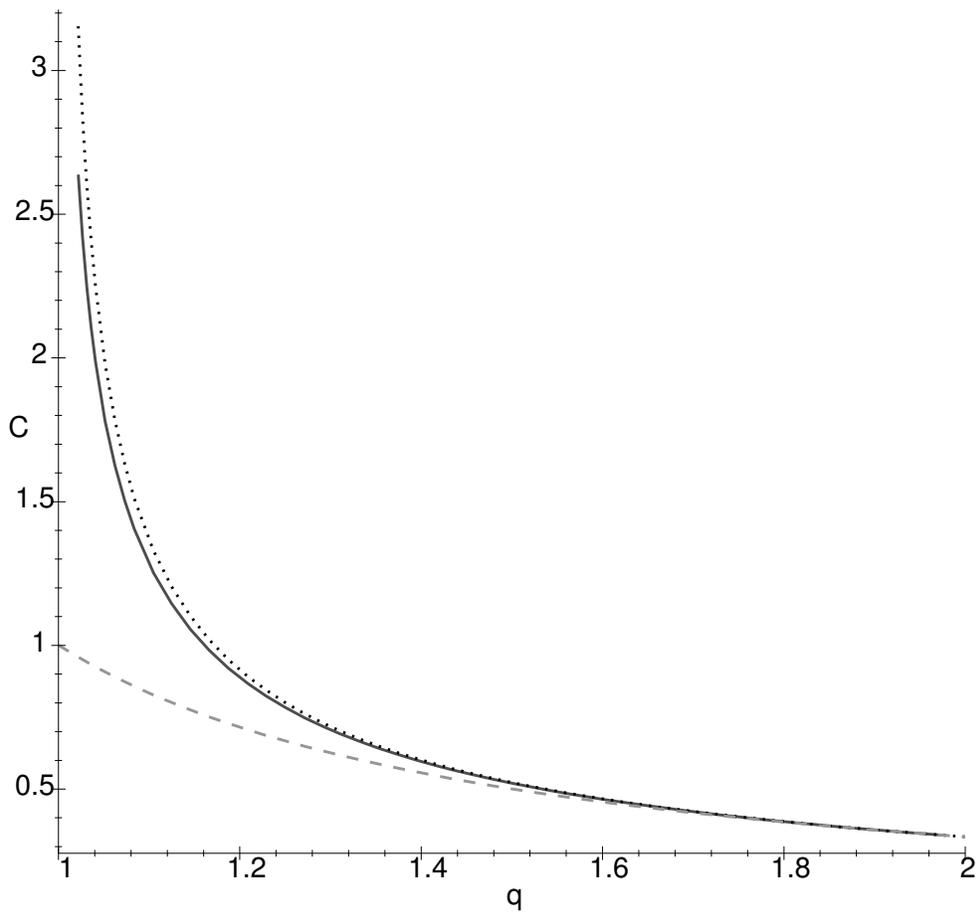}
\caption[]{\label{fig2} $c(Q)$ (solid middle line), $c(O)$ (upper dotted line)
and $c(P)$ (lower dashed line) plotted against $q$ for $1<q<2$.
$c(O)$ and $c(Q)$ diverge as $q \rightarrow 1$.}
\end{figure}
\clearpage \newpage
\begin{figure}[htb]
\vskip 20.0truecm
\includegraphics{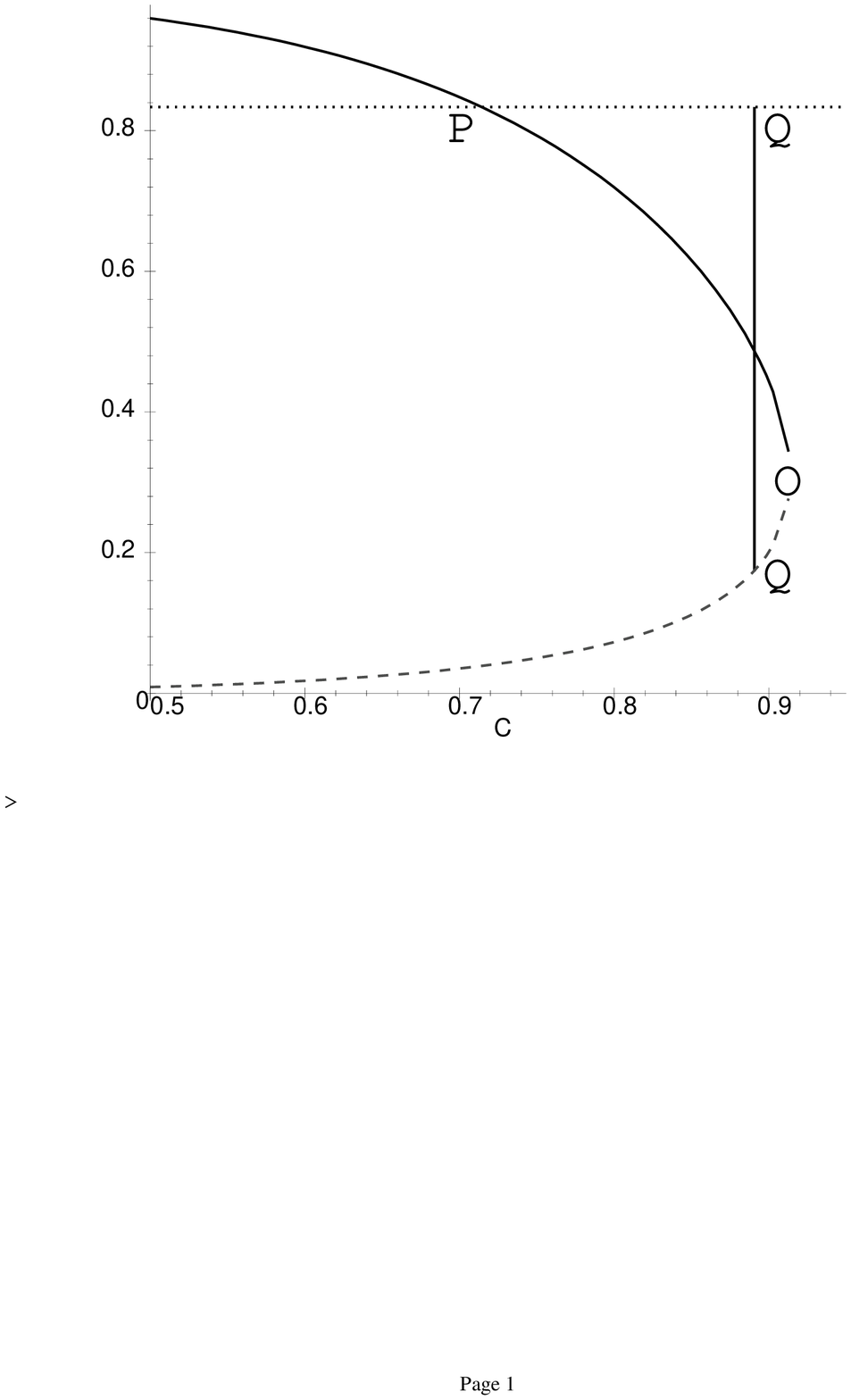}
\caption[]{\label{fig3} The magnetisation $m$
for a $q=1.2$ state Potts model as calculated from
the saddle point solutions. The key is as for Fig.1,
but note that the first order transition at {\bf Q}
is now to the {\it lower} branch.}
\end{figure}

\end{document}